\documentstyle[aps,twocolumn,prl,epsfig,ilja1]{revtex}

\begin{document}

\title{Generation of lattice structures of optical vortices}

\author{Alexander Dreischuh and Sotir Chervenkov}

\address{Department of Quantum Electronics, Sofia University,\\ 
5 James Bourchier Blvd., 1164 Sofia, Bulgaria}

\author{Dragomir Neshev}
\address{Vrije Universiteit Amsterdam, Laser Centre,
De Boelelaan 1081, 1081 HV  Amsterdam, The Netherlands}

\author{Gerhard G. Paulus}
\address{Max-Planck-Institut f\"ur Quantenoptik, Hans-Kopfermann-Str. 1,
D-85748 Garching, Germany}

\author{Herbert Walther}
\address{Max-Planck-Institut f\"ur Quantenoptik, Hans-Kopfermann-Str. 1,
D-85748 Garching, Germany}
\address{Ludwig-Maximilians-Universit\"at M\"unchen, Sektion
Physik, Am Coulombwall 1,  D-85748 Garching, Germany}

\maketitle

\begin{abstract}
We demonstrate experimentally the generation of square and hexagonal
lattices of optical vortices and reveal their propagation in a
saturable nonlinear medium. If the topological charges of the
vortices are of the same sign the lattice exhibit rotation, while if
alternative, we observe stable propagation of the structures. In the
nonlinear medium the lattices induce periodic modulation of the
refractive index. Diffraction of a probe beam by this 
nonlinearity-induced periodic structure is observed.
\end{abstract}


\section{Introduction}
Optical vortices are intriguing objects that attract much attention
\cite{Vasnetsov} and display fascinating properties with possible applications 
in the optical transmission of information, or guiding and trapping of 
particles. They have a characteristic screw-type phase
dislocation \cite{Nye}, which order multiplied
by its sign is referred to as a topological charge (TC). The study of
optical vortices and more general - phase singularities, suggests not only
new directions of fundamental research but also provides links to
other branches of physics, as quantum optics \cite{Furasawa},
superfluidity \cite{Donnelly}, Bose-Einstein condensates
\cite{Williams,Matthews}, and cosmology.

Optical vortices can be generated in several different controllable
ways: in lasers with large Fresnel numbers \cite{Coullet}, by
helical phase plates \cite{Beijersbergen}, laser mode converters
\cite{Tamm,Petrov}, or computer-generated holograms (CGHs)
\cite{Heckenberg}. The method of the CGHs, however, is the most commonly 
used, since it allows precise control of the vortex position, TC, and
possibility for generation of specific patterns of optical vortices.

The propagation dynamics of a single vortex, both in linear and
nonlinear media, has been a subject of many researches
(see, e.g., \cite{Basisty,Freund,Staliunas}), whereby also
the non-canonical properties of the vortex have been taken into account 
\cite{Torner,Gabriel}. It has been
shown that the vortex position on the background beam is strongly
affected by any source of phase and intensity gradients
\cite{Rozas97,Swartzlander97,Kivshar98} and could be controlled by
interference with a weak plane-wave \cite{Christou}. Special attention
attract the vortices propagating in a self-defocusing nonlinear
medium (NLM), where they can form an optical vortex soliton (OVS) 
\cite{dark_review}. (For overview on OVSs see Ref.~\cite{Vasnetsov}, 
chapters 7 - 8.) OVSs induce in the medium optical waveguides
\cite{Snyder,Carlsson,Law2000} which can guide weak information beams. 
An OVS was first experimentally generated in Kerr NLM
\cite{Swartzlander92} and later in media with other types of
nonlinearity: saturable-atomic \cite{Tikhonenko98}, photorefractive
\cite{Duree}, and photovoltaic \cite{Chen}. Recently an OVS was also
observed in a quadratic NLM with defocusing response.
However, attention has been paid to avoid the modulational
instability of the plane-wave background beam \cite{Trapani}.

The propagation of multiple-charged OVSs has also been investigated
\cite{Mamaev97,Velchev}. It was found that 
they are topologically unstable and decay into vortices of unit charge 
\cite{Dreischuh}. The vortices produced by the decay, can arrange themselves 
in regular patterns (vortex ensembles) while interacting with each other by 
phase and intensity gradients. The decay of the higher-order vortices obey
the general principal of conservation the total angular momentum (AM) of
the beam carrying them. Additionally, for a closed region of space the net 
topological charge must be conserved under continuous evolution provided that 
no vortices enter or leave the region. 

An ensemble of optical vortices
exhibits a fluid-like motion \cite{Swartzlander97,Neshev98} which
strongly depends on the geometrical configuration. The propagation of 
the simplest vortex ensemble, namely a vortex pair, has been investigated by 
several groups \cite{Rozas97,Swartzlander97,Neshev98,Barry94,Rozas2000}. 
In Ref.~\cite{Barry94} the rotation of the pair of vortices 
with equal TCs is controlled by the Gyou phase of the host Gaussian beam.
By changing the beam intensity, the position of the beam waist inside the 
self-defocusing NLM changes, thus changing the angle of the rotation at 
the output plane. A comparison between the degree of rotation of a vortex 
pair in linear and nonlinear regime was also performed in 
Ref.~\cite{Rozas2000}. It was pointed out that the effect of rotation in 
the nonlinear regime could be more than three times higher than in the 
linear one. The enhancement is assigned to the nonlinear confinement of
the vortex cores, which allows the vortices to propagate as vortex
filaments.

Recently the propagation of vortex arrays has been investigated. Such 
arrays were generated by a bent glass plate \cite{Kim}, or as a result 
of transverse instability of dark soliton stripes
\cite{McDonald,Tikhonenko96,Mamaev96}. The instability could be enhanced
additionally when the dark-soliton stripe interacts with an
optical vortex, causing ``unzipping'' of the stripe \cite{Neshev2001}.
Ensembles of ordered optical vortices have also been 
investigated in quadratic NLM and are promising for
controllable generation of multiple-vortex patterns \cite{Molina}. The
proposed method paves a way for creation of reconfigurable
vortex ensembles by means of seeded second-harmonic generation.

Regarding the fluid-like motion of the vortex-ensembles, a stationary
configuration of vortices was found \cite{Neshev98}. It consists
of three vortices of equal TC situated in an equilateral triangle and an
additional vortex with opposite TC in the center. That configuration was proven 
to be stable under small displacement of one of the dislocations. However, if 
the vortices are of higher-order they decay and subsequently form 
another stationary configuration resembling a part of hexagonal honey-comb 
lattice. This fact directs attention to the investigation of optical vortex
lattices and characterisation of the propagation of the beams they are 
imposed on. 

Up to now lattices of optical vortices propagating in NLM were
considered only theoretically. The simplest case of square lattice 
consisting of vortices with alternative charges was investigated 
by direct modeling of four of them under periodic boundary conditions 
\cite{Firth}. Later, lattices with different
geometries imposed on a finite background beam (conditions closer to
experimental ones) were considered \cite{Neshev98}. It was shown
that depending on the TCs the vortex lattices can exhibit rotation 
or rigid propagation for equal or alternative TCs, respectively. In 
addition, lattices possess elasticity against displacement
of one or more vortices out of their equilibrium positions.

Here we report the first to our knowledge experimental
investigation of lattice structures of optical vortices in self-defocusing
NLM. We concentrate our attention on two types of lattice geometries - 
square and hexagonal one. When propagating in the NLM they induce a periodic 
modulation of its refractive index. For high beam intensities these changes 
are sufficient to cause a diffraction of a probe beam propagating 
perpendicularly to the volume with periodically modulated refractive index. 
This diffraction may be controlled by steering the propagation of the 
vortex-lattice, e.g. by controlling its degree of rotation (for lattices 
consisting of vortices with equal charges). Additional control may be 
attained by changing the pump beam intensity which changes the refractive 
index of the medium and therefore the diffraction efficiency of the induced 
periodic phase grating.

The maximal refractive index change in our experiment is of the 
order of $10^{-4}$ to $10^{-3}$ which is not enough to form an effective  
two-dimensional photonic band-gap structure \cite{ph_crystal}. As a 
proof-of-principle, however, one can consider the possibility to 
trap glass spheres \cite{Gahagan} by the optical vortices ordered 
in a lattice. This might give an opportunity for generation of effective 
two-dimensional photonic crystals. Such a crystal could
be reconfigured by altering the degree of rotation
of the lattice (by changing the intensity of the focused background beam as 
in Ref.~\cite{Barry94}) for equal TCs, or by use of dynamically reconfigurable 
holograms \cite{holograms}.

We would also like to emphasize the close link between our results
and the field of Bose-Einstein condensates, where
experimental investigations on vortex ensembles \cite{Madison},
vortex arrays as a result of dark soliton-stripe instability 
\cite{Anderson}, and vortex lattices \cite{Ketterle} have been 
reported recently.

\section{General analysis}
Let us consider the propagation of a beam in self-defocusing NLM with saturable
nonlinearity where its evolution is described by the normalized nonlinear 
Schr\"odinger equation (NLSE) for the slowly-varying amplitude envelope

\begin{equation}
\label{Schroedinger}
i\frac{\partial E}{\partial z}+\frac{1}{2}\triangle_{\perp} E 
        - {|E|^2E\over (1+s|E|^2)^\gamma}=0,
\end{equation}
were $\triangle_\perp$ is the transverse Laplace operator. The
transverse coordinates ($x,y$) are normalized to the characteristic 
size of the dark structures $a$, and the propagation coordinate $z$ is
normalized to the diffraction length of the dark beams. The background 
beam intensity $I=|E|^2$ is expressed in units of the intensity necessary 
to form one-dimensional (1D) dark soliton $I_{1Dsol}$ of size $a$.
The saturation parameter is defined by $s=I_{1Dsol}/I_{sat}$, where 
$I_{sat}$ is the saturation intensity retrieved by the experimental
conditions. The model of the saturation we use is introduced 
phenomenologically in order to describe the nonlinear response of the 
thermal medium. It was derived from a test experiment for self-bending of 
the background beam and described in detail in our previous works
\cite{Dreischuh,Dreischuh2000}. The parameters of the nonlinear response 
function $s$ and $\gamma$ depend on the particular realization of the 
experiment, e.g. the properties of the NLM and the focusing conditions. 
In all measurements reported here we use thermal nonlinearity, and in 
particular ethylene-glycol dyed with DODCI (Diethyloxadicarbocyanine 
iodide). Two concentrations of the dye were used so that $s=0.4$ and $1.2$, 
respectively and $\gamma\simeq 3$ in both cases.

In order to investigate the propagation dynamics of vortex lattices 
we first conducted numerical simulations by use of beam propagation method. 
The initial conditions were modelled as superposition 
of vortices situated in the nodes of a lattice:
\begin{equation}
\label{eq2}
E(\vec r,z=0)=\prod_{j,k=-\infty}^\infty \left\{ \matrix{
	\mbox{sq}(\vec r - \vec r_{jk})\cr
	\mbox{hex}(\vec r - \vec r_{jk})} \right.,
\end{equation}
with a square and hexagonal symmetry respectively. In Eq.~(\ref{eq2}) 
$\vec r_{jk}$ are the nodes of the Bravais lattice representing the physical 
lattice structure. The square lattice (Fig.~1 - upper row) coincides with its 
Bravais lattice, however the hexagonal honey-comb lattice (Fig.~1 - second row) 
is represented by as Bravais lattice with a base containing two vortices. If 
one define the primitive vectors of the Bravais lattice as $\vec b$ and 
$\vec c$ then its nodes are described as $\vec r_{jk}=j\vec b+k\vec c$, with 
$j,k$ integer numbers.
The primitive vectors of the square lattice are orthogonal to each other and
can be expressed in $(x,y)$ coordinates as $\vec b= (\Delta,0)$, 
$\vec c= (0,\Delta)$, where $\Delta$ is the distance between two neighbouring
vortices. For the honey-comb lattice the primitive vectors are not
orthogonal and are expressed as: $\vec b= (\sqrt 3\Delta,0)$ and
$\vec c= (\frac{\sqrt 3}{2}\Delta,\frac32\Delta)$. Then the two vortices inside
the elementary cell have positions 
$\vec r_1=\frac13(\vec b+\vec c)$ and $\vec r_2=\frac23(\vec b+\vec c)$.

\vskip -3mm
\begin{figure}[H]
\setlength{\epsfxsize}{3.0in}
\centerline{\epsfbox{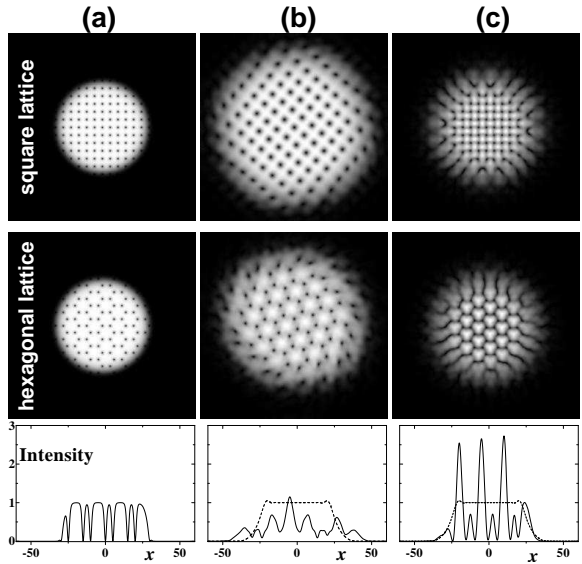}}
\vskip 1mm
\caption{Background beams containing vortex lattices of different
geometries. (a) at the input of the NLM and (b,c) at $z=10$ for
lattice with equal and alternating TCs, respectively. Upper row:
Images of square-shaped lattice; Second row - hexagonal lattice; Bottom row -
transverse slices of hexagonal lattices. For comparison, 
transverse slice of the background beam (without vortices nested in) 
is shown with dashed curve. In all cases $\Delta=5.0$, $I_0=1$, and $s=0.4$. 
The images are grayscale-coded and white corresponds to the maximal intensity.}
\end{figure}

The functions $\mbox{sq}(\vec r - \vec r_{jk})$ and 
$\mbox{hex}(\vec r - \vec r_{jk})$ describe the structure of the elementary
cell of the Bravais lattice and are expressed as:
\begin{equation}
\label{eq3}
\mbox{sq}(\vec r - \vec r_{jk})=\tanh(|\vec r - \vec r_{jk}|)
\exp\left[i\;\mbox{sgn}^{j+k}\arctan\frac{(\vec r - \vec r_{jk})_y}
	{(\vec r - \vec r_{jk})_x}\right],
\end{equation}
and
\begin{eqnarray}
\label{eq4}
\nonumber
\mbox{hex}(\vec r - \vec r_{jk})=
	\tanh(|\vec r - \vec r_{jk} - \vec r_1|) \;
	\tanh(|\vec r - \vec r_{jk} - \vec r_2|) \\
\times\;\exp\left[i\;\arctan\frac{(\vec r - \vec r_{jk} - \vec r_1)_y}
	{(\vec r - \vec r_{jk}- \vec r_1)_x}\right]
\exp\left[i\;\mbox{sgn}\arctan\frac{(\vec r - \vec r_{jk} - \vec r_2)_y}
        {(\vec r - \vec r_{jk}- \vec r_2)_x}\right].
\end{eqnarray}
The sign function $\mbox{sgn}$ is equal to $+ 1$ for equal TCs and $-1$ for 
alternative ones. 

Then the lattice structure is imposed on a super-Gaussian (flat-top) 
background beam
\begin{equation}
\label{Super-Gaussian}
B(x,y,z=0)=\sqrt{I_0}
\exp\left\{-\left(\sqrt{\frac{x^2+y^2}{w^2}}\right)^{14}\right\},
\end{equation}
where the width $w$ is chosen to exceed the characteristic width of the 
dark structures $a$ more than $40$ times, and $I_0$ is the maximal background
beam intensity.  

We modelled the propagation of lattices of different geometries and different 
TC distributions (see Fig.~1). No qualitative differences 
were observed in the propagation of vortex-structures with respect to
the lattice geometry (square or hexagonal). The mode of propagation, however, 
crucially depends on the vortex charge distribution (equal - Fig.~1b or 
alternative - Fig.~1c). Two characteristic differences are clearly seen: 
(i) In the case of equal TCs ($\mbox{sgn}=+1$) the superposition of the 
phases of all vortices results in azimuthal phase gradient and nonzero 
total AM which causes rotation of the whole structure (Fig.~1b). In the 
case of alternative TCs ($\mbox{sgn}=-1$) the superposition of all the 
phases gives, in average, no phase gradient and zero total AM. As a result, 
steady propagation of the lattice is observed in the simulations (Fig.~1c);
(ii) In the case of equal TCs the non-zero total AM and the centrifugal forces
lead to increased broadening of the background beam. The maximal intensity 
rapidly decreases along the NLM ($I\simeq 0.6$ at $z=10$ Fig.~1b - bottom row). 
The dependence on the intensity of the background beam in this case is 
relatively weak and the topological effects dominate the nonlinear ones.
In the case of alternative TCs (Fig.~1c) the background beam broadening is 
an effect only due to the combined action of the diffraction and 
self-defocusing nonlinearity and depends on the intensity. 

The degree of rotation of both lattice geometries for the case of equal
TCs is depicted in Fig.~2. The rotation is due to the phase
gradient, which is higher for the denser (square-shaped) structure. Therefore
its rotation (open squares in Fig.~2) is faster than the rotation of the 
hexagonal one (solid circles). The dependence on the distance is not linear 
because in the course of propagation the background beam spreads out and the 
distance between the vortices increases. That causes a decrease of the angular 
velocity with the propagation length.

\vskip 1mm
\begin{figure}[H]
\setlength{\epsfxsize}{2.3in}
\centerline{\epsfbox{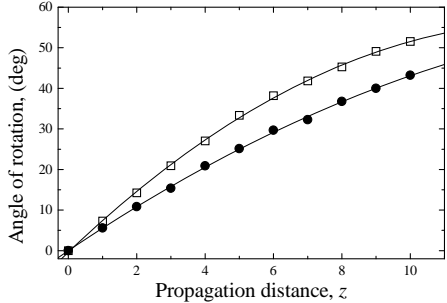}}
\vskip 1mm
\caption{Angle of rotation of vortex lattices of equal TCs
vs. propagation distance for square (open squares) and hexagonal
(solid circles) geometry. The solid lines are drown to
guide the eye. The lattice parameters are the same as in Fig.~1.}
\end{figure}

Similar behaviour was previously described for the case of Kerr
nonlinearity \cite{Neshev98}. Here we would like to point out the
features caused by the saturation of the nonlinearity. As already mentioned, 
the rotation of the lattice and the increased beam spreading in the case 
of equal TCs are topological effects. The effects which depend on the
nonlinearity are related to the local intensity via the specific beam shape 
and the vortex-pattern formed on the background. For example, the transverse
profile of a single OVS in saturable medium differs significantly at different 
values of the saturation parameter \cite{Tikhonenko98} (the OVS is broader at
higher saturation). In the case of periodically ordered vortices when
each individual dark beam starts to broaden, the overlapping with the wings 
of its neighbours increases, while the individual cores do not change 
significantly. Since the vortices are imposed on a finite background beam 
and its total energy is conserved, bright peaks form in between as a result 
of local energy redistribution (See Fig.~1c - bottom row). Therefore, even 
in saturable medium the well defined periodic modulation of the 
refractive index of the medium is still preserved.

\section{Experimental investigation}
The experimental setup is similar to the one used in our previous
works \cite{Dreischuh,Dreischuh2000} and is shown in Fig.~3. The
$488nm$ line from an $Ar^+$ laser is used to reconstruct the 
photolithographically produced CGH with the desired vortex-lattice. 
The $+1$ (or $-1$) order of the diffraction is separated by an iris 
diaphragm (D) and is focused on the input face of a glass cell containing 
the NLM. The output face of the cell is imaged to a CCD camera, and 
neutral filters (F) are used to avoid its saturation.

\vskip -3mm
\begin{figure}[H]
\setlength{\epsfxsize}{3.0in}
\centerline{\epsfbox{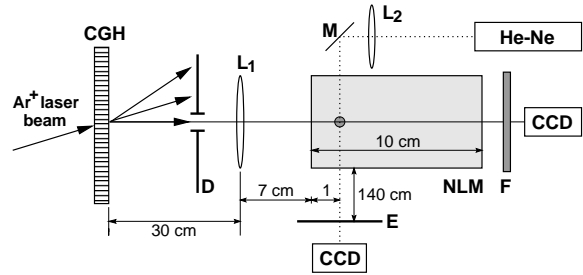}}
\vskip 1mm
\caption{Experimental setup: CGH - computer generated hologram; D -
iris diaphragm; L$_1$, L$_2$ lenses of focal length $7.0 cm$ and $8.0cm$,
respectively; M - mirror; E - screen; NLM - nonlinear medium; F - neutral 
filters; CCD - camera. The characteristic distances are shown.}
\end{figure}

To assure the correct generation of the lattices by the CGHs we first
opened the diaphragm allowing the $+1$ diffraction order to
interfere with the plane zeroth one. Interference patterns for three
different cases are presented in Fig.~4. The vortices appear as forks
of interference lines. Two neighbouring vortices at each image are
marked with arrows. The images show correctly reproduced square-shaped
lattice with alternative TCs and two hexagonal lattices with
alternative and equal TCs, respectively (Fig.~4, left to right). The
images are brighter in the right-hand side since they overlap
inhomogeneously with the zeroth order beam. That inhomogeneity also
introduces an intensity gradient in the structure of vortices, which 
causes shrinking and vortex displacement from their regular positions. 
In overall that leads to deformation of the lattice. Being aware of this 
fact in the experiment we preserved the regular lattice structure by
placing the diaphragm as close as possible to the CGH.

\vskip -3mm
\begin{figure}[H]
\setlength{\epsfxsize}{3.0in}
\centerline{\epsfbox{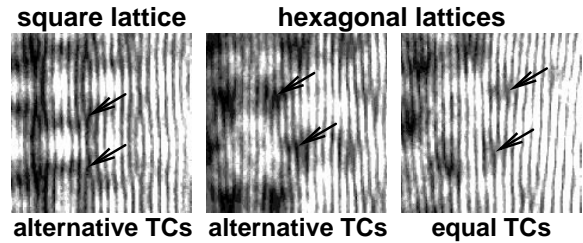}}
\vskip 1mm
\caption{Interferograms of three experimentally generated lattices.
From left to right: square lattice with alternative TCs; hexagonal
lattice with alternative TCs; and hexagonal lattice with equal
TCs. Two neighbouring vortices in each image are marked with arrows.}
\end{figure}

The features of the nonlinear propagation are determined by measuring the 
characteristic nonlinear parameters of the media for the two concentrations 
of the dye. For the lower one, the power necessary to form a 1D 
dark-soliton stripe was estimated to be $P_{1Dsol}\simeq 22mW$ and the 
saturation power was $P_{sat}\simeq 60mW$ (measured in a self-bending 
scheme) \cite{Dreischuh2000}. For the higher concentration the 
characteristic powers were $P_{1Dsol}\simeq 20mW$ and $P_{sat}\simeq 16mW$. 
The intensity distributions for two hexagonal lattices (with alternative 
and equal charges) at the end of the NLM with lower dye concentration are 
shown in Fig.~5. Because of some technical restrictions in 
synthesizing the CGH, for the lattice with equal TCs, the number of vortices 
encoded is less than in the hologram with alternative TCs. The 
geometrical characteristics (the elementary cell of the lattices), however, 
are the same in both cases. The propagation behaviour for both lattices 
is clearly different. While the lattice with alternative charges exhibits 
steady propagation (Fig.~5a) the one with equal ones (Fig.~5b) tends to rotate
(at about $28^\circ$ counter-clockwise). The background beam spreads stronger 
than in the case of a lattice with alternative charges. Unfortunately, due 
to the different number of vortices, this fact is not obvious in Fig.~5. 
The smaller number of vortices in Fig.~5b modulates the background beam in 
a way that more filters were used to avoid the saturation of the CCD camera. 
As a consequence, the wings of the background beam in Fig.~5b are not seen 
and the beam diameter seems to be smaller as compared to Fig.~5a. In order to 
illustrate that the spreading is indeed higher in the case of equal TCs we 
looked in detail on the size of the elementary cell of the honey-comb lattice. 
Since the distances between the neighbouring vortices were encoded in the 
CGHs to be the same (the produced holograms were inspected by a microscope) 
any difference in the vortex separation is due to the evolution during 
propagation. In each image in Fig.~5 we inset the exact size and
orientation of the elementary hexagonal cell of the lattice (see
bottom-right corner of each image). Indeed the comparison between the
elementary cells of the lattices in both cases shows $18\%$ bigger 
size for the one with equal TCs.

\vskip -3mm
\begin{figure}[H]
\setlength{\epsfxsize}{2.4in}
\centerline{\epsfbox{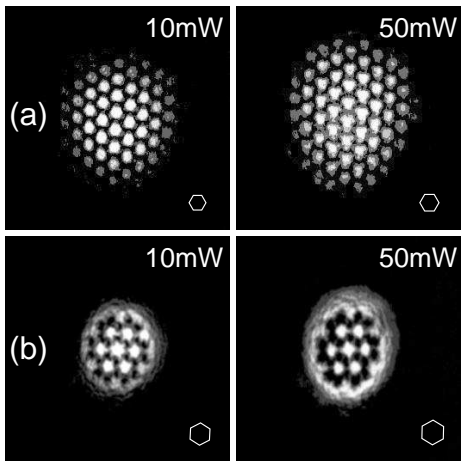}}
\vskip 1mm
\caption{Experimental images of the vortex lattices after $10 cm$
propagation in the NLM. (a) - hexagonal lattice with alternating TCs for
powers of $10mW$ and $50mW$. (b) - hexagonal lattice with equal TCs 
for the same powers. The insets in each image represent the 
size and the orientation of the elementary cell of each lattice.}
\end{figure}

The influence of the nonlinearity can be seen if the corresponding images 
for two particular powers are compared. We note again that the increase of 
the beam power does not influence the degree of rotation of the lattice 
presented in Fig.~5b since the waist of the laser beam is near the input face 
of the NLM. The higher power of the laser beam contributes, however, to the 
background beam broadening. Comparing the size of the elementary cell of 
the same lattice at two different powers we estimated $15\%$ broadening of the 
beam for the case in Fig.~5a and $12\%$ for the case in Fig.~5b. That difference 
we attribute to the increased background beam size at the entrance of the 
NLM, which is due to the topological interaction of the equally charged 
vortices between the CGH and the NLM. 

The square-shaped lattices were investigated in the same way and 
qualitatively similar features were observed. We also investigated lattices 
with intentionally encoded defects in their structure, e.g. when one of the 
vortices is missing or all of the vortices in a line are shifted out of their 
equilibrium position. These experiments revealed the interesting property 
that the lattices exhibit elasticity. However the resolution in our
experiments was not sufficient to resolve this feature in more than
qualitative manner.

\section{Diffraction of a probe beam by vortex lattices}
When an intense laser beam propagates along a NLM its refractive index changes
proportionally to the intensity distribution. Since the 
vortex-lattices posses a periodic intensity distribution (see Fig.~1 and 
Fig.~5) one can expect periodic modulation of the refractive index. 
In a self-defocusing medium the higher local intensity will correspond 
to lower local refractive index. The lattices are imposed on a 
finite background beam and as whole it induces in the NLM
a cylindrical defocusing lens (considered in a perpendicular direction), 
which is modulated by the dark vortex structure composing the lattice. 
In a thermal NLM, as slightly absorbing liquid, the nonlocal effect coming 
from the heat transfer also influence the refractive index change and 
effectively decreases its modulation. The nonlocality is not taken into 
account in the model of Eq.~(\ref{Schroedinger}). Its main influence 
is that at zero intensity (the points of vortex-phase dislocation) the
refractive index change is non-zero (see a description in 
Ref.~\cite{Deykoon}).

To investigate the modulation of the refractive index in the NLM caused by 
the presence of lattices we conducted an experiment in which a (probe) 
single-mode He-Ne laser beam was directed perpendicularly to then (pump) 
Ar$^+$ laser beam, as shown in Fig.~3. We aligned the probe beam in a way 
to cross the pump one $1cm$ inside the NLM. (The higher dye concentration 
was used in this experiment). The input profile of the He-Ne laser beam is 
shown in Fig.~6a and its circular symmetry is evident. When it crosses the 
Ar$^+$ laser beam the symmetry is distorted and the beam elongates in 
direction perpendicular to the plane of Fig.~3. 

\vskip -3mm
\begin{figure}[H]
\setlength{\epsfxsize}{3.0in}
\centerline{\epsfbox{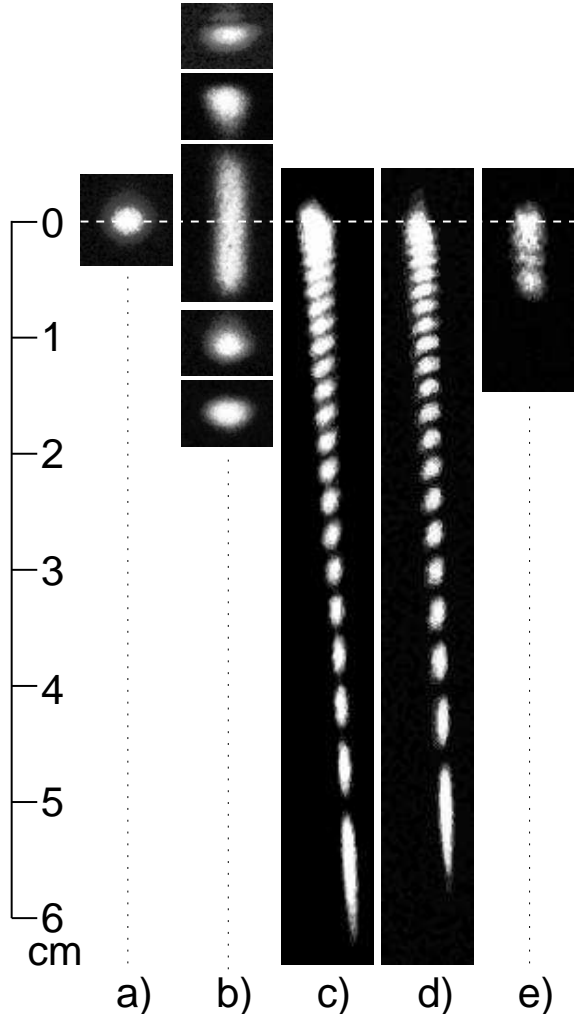}}
\vskip 2mm
\caption{Images of the probe He-Ne beam on the screen E. (a) the input
He-Ne laser beam; (b) the intensity profile of the probe beam at low pump power 
($10mW$) for different parallel vertical displacements with respect 
to the pump one; (c), (d) diffraction of the He-Ne beam from the periodic phase 
gratings induced in the NLM by square and hexagonal lattices, respectively
(pump power $80mW$); (e) the diffraction pattern when a
single vortex is imposed on the pump beam (pump power $30mW$).}
\end{figure}

First we identified the effect of optically-induced Gaussian cylindrical 
lens on the probe beam. In order to keep the same background beam 
characteristics the CGH was shifted in a way that only a region with 
parallel interference lines was illuminated thus ensuring unmodulated 
background beam. This unmodulated pump beam induces a cylindrical lens in 
the NLM whereas the probe one passes trough the lens and defocuses 
(Fig.~6b - central image). The diameters of both beams were estimated to 
be approximately equal at the cross point. Therefore one should expect 
that the probe beam will be strongly affected by aberrations of the 
induced lens. In Fig.~6b five probe beam profiles are shown for different 
positions of the He-Ne laser beam. The pump power is kept $10mW$. Different 
input positions of the probe beam are achieved by parallel vertical 
translation by a simple periscopic system denoted for simplicity as mirror 
M in Fig.~3. The He-Ne beam elongates symmetrically if it crosses the pump 
in the center and asymmetrically if it is shifted up or down. At higher 
powers the aberration of the induced cylindrical lens becomes vertically 
asymmetric, probably due to the asymmetric heat diffusion in the cell. 

The situation is different when the vortex-lattice is imposed on the 
background beam. At power of the Ar$^+$ laser beam higher than $20mW$ the 
vortices have well confined cores. Due to the nonlinear change of the 
refractive index the vortex lattice `writes' a phase grating in the NLM.
The perpendicularly propagating He-Ne laser beam passes trough this grating
and develops well pronounced diffraction orders at the output screen as seen 
in Fig.~6c,d at $80mW$ pump power. The constant of the phase grating written 
is apparently different for the square-shaped lattice (Fig.~6c) and for the
hexagonal-shaped one (Fig.~6d). In the first case the period of the 
vortex-structure is smaller (so this of the phase grating) and higher 
dispersion in the diffraction orders is observed (higher angle of 
diffraction). At lower powers diffraction orders were also observed.
However, it was more difficult to distinguish them at the screen (E)
because the effective cylindrical lens had larger focal length. At different 
powers the magnitude of the refractive index changes and the modulation 
depth of the phase grating written in the NLM is different. This
influences the energy redistribution between the diffraction orders. 
Moreover, because of the finite number of the vortices in
the lattices and the nearly equal sizes of the pump and the probe 
beams the diffraction from the phase grating couldn't be compared directly 
with the diffraction from an infinite periodic structure. In our opinion the
ratio between the intensities of the different diffraction orders is
gradually influenced by the fact that different parts of the probe beam pass
trough different number of vortices. Further, at the exit of the phase 
grating, the modulated probe beam is additionally affected by the aberration 
of the thermal lens.

To assure that the observed diffraction structure is really induced by
the periodicity of the vortex lattices we tested the diffraction
caused by a single vortex imposed on the background beam. As seen in
Fig.~6e the diffraction by a single vortex is substantially different 
and resembles the diffraction of a laser beam by a wire.
In all our experiments we observed strong vertical asymmetry of the
probe-beam diffraction pattern, which always developed downwards at 
higher powers. Numerical modeling of the processes and further experimental 
investigations should allow to gain deeper insight in the relative strength 
of the mentioned mechanisms.

\section{Conclusion}
In conclusion, we successfully generated experimentally lattice
structures of optical vortices with different topological charge distribution
and described their propagation in saturable NLM. Due to the intensity 
dependence of the refractive index these lattices induce periodic modulation
of the refractive index of the medium and `write' and effective phase grating
in it. The modulation is sufficient to force a perpendicularly propagating 
probe beam of a He-Ne laser to diffract.
This feature could appear as an interesting possibility to 
create periodic structures in the refractive index of a NLM. It could
find an application for optical writing of two-dimensional photonic 
crystals and could appear relevant to the physics of Bose-Einstein condensates. 

\section{Acknowledgements}
A.D. thanks the Alexander-von-Humboldt foundation for the
awarded fellowship and for the possibility to carry out the
measurements at the Max-Planck-Institut f\"ur Quantenoptik
(Garching, Germany). The work of D.N. was partially supported by
Marie-Curie Individual Fellowship under contract HPMFCT-2000-00455. 
The authors thank to Yu. Kivshar, L. Torner, A. Desyatnikov, and 
N. Herschbach for the valuable discussions and the support of this work.


\end{document}